# Temporal and spectral fingerprint of ultrafast all-coherent spin switching

S. Schlauderer[1,*], C. Lange[1,*,†], S. Baierl[1], T. Ebnet[1], C. P. Schmid[1], D. C. Valovcin[2], A. K. Zvezdin[3,4], A. V. Kimel[5,6], R. V. Mikhaylovskiy[6,†,§] and R. Huber[1]

[1]Department of Physics, University of Regensburg, Regensburg 93053, Germany.
[2]Department of Physics and the Institute for Terahertz Science and Technology, University of California at Santa Barbara, Santa Barbara, California 93106, USA
[3]Prokhorov General Physics Institute and P.N. Lebedev Physical Institute of the Russian Academy of Sciences, Moscow 119991, Russia.
[4]Moscow Institute of Physics and Technology (State University), Dolgoprudny 141700, Russia.
[5]Moscow Technological University (MIREA), Moscow 119454, Russia.
[6]Radboud University, Institute for Molecules and Materials, Nijmegen 6525 AJ, The Netherlands.

**Future information technology demands ultimately fast, low-loss quantum control. Intense light fields have facilitated important milestones, such as inducing novel states of matter[1-3], accelerating electrons ballistically[4-7], or coherently flipping the valley pseudospin[8]. These dynamics leave unique signatures, such as characteristic bandgaps or high-order harmonic radiation. The fastest and least dissipative way of switching the technologically most important quantum attribute – the spin – between two states separated by a potential barrier is to trigger an all-coherent precession. Pioneering experiments and theory with picosecond electric and magnetic fields have suggested this possibility[9-11], yet observing the actual dynamics has remained out of reach. Here, we show that terahertz (1 THz = $10^{12}$ Hz) electromagnetic pulses allow coherent navigation of spins over a potential barrier and we reveal the corresponding temporal and spectral fingerprints. This goal is achieved by coupling spins in antiferromagnetic $TmFeO_3$ with the locally enhanced THz electric field of custom-tailored antennas. Within their duration of 1 ps, the intense THz pulses abruptly change the magnetic anisotropy and trigger a large-amplitude ballistic spin motion. A characteristic phase flip, an asymmetric splitting of the magnon resonance, and a long-lived offset of the Faraday signal are hallmarks of coherent spin switching into adjacent potential minima, in agreement with a numerical simulation. The switchable spin states can be selected by an external magnetic bias. The low dissipation and the antenna's sub-wavelength spatial definition could facilitate scalable spin devices operating at THz rates.**

The lowest theoretical limit of energy dissipation for manipulating one bit of information is defined by the Landauer principle[12] as $Q = k_\text{B}T \ln 2$, where $T$ is the temperature and $k_\text{B}$ denotes the Boltzmann constant. This can be seen as a result of inelastic scattering of a quasiparticle of energy $Q$, such as a collective spin excitation, called a magnon. At or below room temperature, $Q$ is of the order of meV, which by the uncertainty principle entails picosecond time scales for minimally dissipative switching. Thus, precessional switching[10,13,14] triggered by a single-cycle THz pulse with meV photon energies and sub-picosecond duration promises ultimately fast and least-dissipative magnetic control.

Experimentally, ultrafast spin control has come a long way[15-17] from the discovery of subpicosecond laser-induced spin dynamics[18] to all-optical non-thermal recording[19]. Understanding strongly non-equilibrium spin dynamics triggered by THz pulses, however, is still in its infancy. In antiferromagnets, magnons feature resonance energies in the meV range[20] and can be directly addressed by the magnetic field component of intense THz pulses[21-23]. Since the underlying Zeeman interaction is relatively weak, magnetic field amplitudes, which allow for a complete spin reversal have only been reached in linear accelerators[9], where the spin dynamics have not been accessible on the intrinsic femtosecond scale. Also spin transfer torques mediated by THz-driven electric currents have induced switching of antiferromagnetic domains, yet without ultrafast temporal resolution[24].

Conversely, electromagnons and the more universal coupling of crystal-field split electronic transitions or coherent phonons with the magnetic anisotropy field have allowed the electric THz field component to drive large-amplitude magnons, observed directly in the time domain[22,25,26]. The available THz peak electric field of 1 MV cm$^{-1}$, however, has limited the maximum spin excursion far below critical values needed for a complete spin reversal. Meanwhile, the near-field enhancement in custom-tailored antenna structures has been exploited to sculpt atomically strong THz waveforms, sufficient to drive non-perturbative nonlinearities, such as THz-induced phase transitions[27] and interband Zener tunnelling, with subdiffractional spatial definition[28]. Such enhancement of the electric field has not yet been utilized for coherent spin control.

Here we combine the advantages of electric-field induced anisotropy changes in an antiferromagnet with the local near-field enhancement of metal antennas. We ballistically steer spins over potential barriers to achieve THz-driven switching between stable states while these dynamics are observed directly on the femtosecond scale. The experiments are performed in high-quality single crystals of the model antiferromagnet $TmFeO_3$. The antiferromagnetically ordered $Fe^{3+}$ spins are slightly canted by the Dzyaloshinskii-Moriya interaction, resulting in a net ferromagnetic moment. As the magnetic anisotropy depends on temperature[26], the spins undergo reorientation phase transitions at $T_1$ = 80 K and $T_2$ = 90 K. The anisotropy may also be modified by THz electric dipole transitions between crystal field-split states of the electronic ground state of the $Tm^{3+}$ ions, the angular momenta of which are coupled with the $Fe^{3+}$ spins by exchange and dipolar interactions[29]. Our idea is to abruptly change the magnetic anisotropy by sufficiently strong THz pulses causing the spins to switch fully ballistically.

We fabricate custom-tailored bowtie antennas of gold (feed gap, 3.5 µm) onto a 60-µm-thick single crystal of $TmFeO_3$ (Extended Data Figure 1) to bypass the diffraction limit and maximize the achievable THz amplitude. The design was guided by numerical finite-difference frequency-domain simulations optimizing the near-field enhancement at a frequency of 0.65 THz (see Methods), which is resonant with crystal field-split ground state transitions in $Tm^{3+}$. In a pump-probe scheme (Fig. 1a), an intense THz transient with tuneable far-field amplitudes of up to $E_{THz}$ = 1.0 MV cm$^{-1}$ (see Methods) excites the structure from the $TmFeO_3$ back side. The ensuing spin dynamics are probed via the polarisation rotation, $\theta$, imprinted on a co-propagating femtosecond near-infrared pulse (wavelength, 807 nm; pulse duration, 33 fs) by the Faraday effect and magnetic linear dichroism. Our quantitative simulation shows that, for the strongest electro-optically detected THz waveform, the near-field of the antenna, $E_{NF}$, readily exceeds 9 MV cm$^{-1}$ in the centre of the gap (Fig. 1b).

To test the efficiency of the antenna, we compare the magneto-optical signal induced in $TmFeO_3$ in the transition phase ($T$ = 83 K) with and without the near-field antenna, as a function of the pump-probe delay time, $t$. In the absence of an antenna, a THz pulse with an amplitude of $E_{THz}$ = 1.0 MV cm$^{-1}$ abruptly sets off coherent magnon oscillations, which decay exponentially within 40 ps (Fig. 1c, black curve). The signal consists of a superposition of two frequency components centred at 0.09 THz and 0.82 THz (inset of Fig. 1c) – the quasi-ferromagnetic (q-fm) and the quasi-antiferromagnetic (q-afm) mode[26], respectively. The maximum rotation angle of the probe polarisation of 0.5 mrad corresponds to a magnetisation deflection by 3.5° (see Methods). In contrast, we observe a qualitatively different response when the probe pulse is positioned in the centre of the antenna feed gap. Here a polarisation rotation as high as 0.9 mrad is reached for a much weaker THz far-field of $E_{THz}$ = 0.4 MV cm$^{-1}$ (Fig. 1c, blue curve). In addition, the relative spectral amplitude of the q-fm mode is significantly enhanced, whereas the amplitude of the q-afm mode is suppressed. This behaviour is expected since the q-fm mode is excited by the antenna-enhanced THz electric field component, whereas the q-afm magnon can only be launched by Zeeman coupling to the THz magnetic field[26], which is not enhanced in the feed gap.

The amplitude of the q-fm magnon is remarkably high given that the field enhancement is spatially confined to the evanescent near-field region (depth, ~13 µm) whereas the magneto-optical signal in the antenna-free case (Fig. 1c, black curve) originates from the entire thickness (60 µm) of the TmFeO$_3$ substrate. A rough estimate (see Methods) shows that the spins in the antenna gap need to undergo a rotation by as much as 24° in order to explain the observed signal strength. Hence, a further increase of the incident THz field may be able to cause complete spin switching.

Figure 2a shows the ultrafast polarisation rotation probed in the feed gap, for various far-field THz amplitudes between $E_{THz}$ = 0.15 MV cm$^{-1}$ and 1.0 MV cm$^{-1}$. For the lowest field, the spin dynamics resembles the q-fm precession sampled in the unstructured crystal (Fig. 1c, black curve). For increasing fields, the oscillation amplitude grows. When the incident THz field exceeds $E_{THz}$ = 0.75 MV cm$^{-1}$, a qualitatively new behaviour emerges. The period of the first full cycle of the magnetisation oscillation is distinctly stretched (see vertical dashed line in Fig. 2a) while a pronounced beating feature occurs in the coherent polarisation rotation signal, seen during 25 ps < $t$ < 35 ps. Simultaneously, a long-lived offset of the Faraday signal develops (Fig. 2b, red shaded area). In the frequency domain (Fig. 2c), these novel dynamics are associated with an asymmetric splitting of the q-fm magnon resonance superimposed on a broad spectral distribution, somewhat reminiscent of the spectral fingerprint of carrier-wave Rabi oscillations[30]. The long-lived offset (Fig. 2b) manifests itself in a dc spectral component, $A_{0\,THz}$, which grows more rapidly for $E_{THz}$ > 0.75 MV cm$^{-1}$ (Fig. 2d and Extended Data Figure 2). We will show next that the stretching of the first oscillation cycle, the beating of the Faraday signal, and the spectral splitting of the magnon resonance are hallmarks of all-coherent non-perturbative spin trajectories between adjacent minima of the magnetic potential energy, whereas the long-lived offset directly reads out the switched spin state.

The dynamics can best be understood by starting out with the magnetic structure of TmFeO$_3$ (Fig. 3a). The slight canting between the magnetisations $\mathbf{M}_1$ and $\mathbf{M}_2$ of the two antiferromagnetic sublattices causes a weak ferromagnetic moment $\mathbf{F} = \mathbf{M}_1 + \mathbf{M}_2$ in the *x*-*z*-plane. The antiferromagnetic vector $\mathbf{G} = \mathbf{M}_1 - \mathbf{M}_2$ encloses an angle $\phi$ with the *x*-axis. In the $\Gamma_{24}$ transition phase ($T_1 < T < T_2$), $\phi$ shifts continuously between 0° and 90° as the magnetic potential $W(\phi)$ changes with the thermal population of the Tm$^{3+}$ crystal field-split states[26]. $W(\phi)$ features four intrinsically degenerate minima. To ensure that the pump-probe experiment starts with the same equilibrium spin orientation angle $\phi_0$ for every laser shot, we apply a weak external magnetic field $B_{ext}$ = 100 mT (see Methods).

When the intense THz near-field excites the Tm$^{3+}$ ions, it abruptly modifies $W(\phi)$, shifting both the position, $\phi_0$, and the depth of the potential minimum (Fig. 3b, inset). These non-adiabatic changes give rise to a displacive and an impulsive anisotropy torque, which initiate coherent magnetisation dynamics as described by the generalized sine-Gordon equation (see Methods). Figure 3b illustrates two typical spin trajectories. For a peak near-field of $E_{NF}$ = 6 MV cm$^{-1}$, the spins carry out a coherent oscillation about $\phi_0$. A field of $E_{NF}$ = 10 MV cm$^{-1}$, in contrast, allows the spins to overcome the potential barrier at $t$ = 3.4 ps, and relax into a new equilibrium position $\phi_1$, corresponding to a spin rotation by ~90°. While crossing the potential maximum the spins acquire a characteristic phase, which causes a retardation by ~180° with respect to spin oscillations in the initial potential minimum, seen at $t$ = 9.7 ps (Fig. 3b, red solid line). Once the spins have reached their maximum positive deflection they oscillate back, but do not overcome the potential barrier a second time because of damping. They rather stay within the new minimum and, in a strongly anharmonic motion, accumulate more phase retardation such that the red and blue trajectories in Fig. 3b oscillate in phase again, around $t \approx$ 25 ps.

To link these dynamics with the measured polarisation rotation, we calculate the expected Faraday signal by projecting the ferromagnetic moment $\mathbf{F}(\phi)$ onto the wave vector of the near-infrared probe pulse, $\mathbf{k}_{NIR}$ (see Fig. 3a). By superimposing the contributions of the two spin trajectories in Fig. 3b, the pronounced beating feature ($t \approx$ 25 ps) can be associated with the phase slip occurring during spin switching (see Extended Data Figure 3). For a quantitative analysis, we combine our calculation of the near-field induced by the experimental THz wave with a numerical solution of the local generalized sine-Gordon equation (see Methods). We then weigh the locally induced Faraday signal by the intensity distribution of the probe beam and sum all the microscopic contributions from the probed volume. Figure 3c shows the calculated polarisation rotations, $\theta$, for $E_{THz}$ = 0.4 MV cm$^{-1}$ and 1.0 MV cm$^{-1}$ (fit parameters, see Methods). All experimental features are quantitatively reproduced, including the quasi-monochromatic magnon oscillation, for low fields (Fig. 3c, blue curve), as well as the phase retardation of the first magnon oscillation period and the pronounced beating at $t \approx 25$ ps, at large THz fields (Fig. 3c, red curve). Moreover, the model unambiguously connects the asymmetric splitting of the q-fm resonance and the broad low-frequency components (Fig. 3d) to THz-driven all-coherent spin switching.

The calculation also confirms that the switched spins can be directly read out. As seen in Fig. 2d, increasing $E_{THz}$ leads to a long-lived signal offset. This is caused by two distinct mechanisms: (i) THz excitation of Tm$^{3+}$ ions slightly shifts the position of the potential minimum (Fig. 3b, inset). (ii) A transfer of spins over the barrier can also change the net magneto-optical signal if $\mathbf{k}_{NIR}$ is tilted out of the $y$-$z$-plane (Fig. 3a). Assuming a tilt angle of 1.25°, we can fit all measured transients in the time domain (Fig. 2). For the corresponding fit parameters, $A_{0\,THz}$ traces the experimental field scaling (Fig. 2d), including the slow increase below the switching threshold, $E_{THz} > 0.75$ MV cm$^{-1}$, and the steep slope above (Fig. 3e, red spheres). In contrast, a calculation with identical fit parameters but a tilt angle of 0° yields a slow increase of $A_{0\,THz}$ for all field strengths (Fig. 3e, red circles). From this comparison we assign the slow increase to the shift of the potential minimum whereas the steep slope observed for finite tilt angles can be directly related to the spin transfer over the barrier.

Based on the microscopic understanding of the spin dynamics, we can shape the spin trajectory by tailoring the magnetic potential. As a first control parameter (see Extended Data Figure 4), we lower the temperature to $T = 82.5$ K, where the barrier height, $w$, is slightly increased (Fig. 4a). Consequently, the switching dynamics are decelerated and the beating signature is delayed to $t = 45$ ps (Fig. 4e, top curve). Meanwhile, the spectrum remains qualitatively similar (Fig. 4f, top curve). The barrier height can also be raised by rotating the external magnetic bias field, $\mathbf{B}_{ext}$, by an angle $\alpha = 15°$ about the optical axis (Fig. 4b and Extended Data Figure 1), resulting in a shift of the beating feature to a delay of $t = 55$ ps (Fig. 4e). Thereby, the potential shoulder at $\phi = -115°$ is lowered (Fig. 4b), which enables large-amplitude oscillations throughout a slightly wider potential trough, causing a weak red-shift of the spectrum (Fig. 4f). For α = 60° (Fig. 4c), the dynamics are strongly altered (Figs. 4e, f). After the spins are driven up the potential barrier at $\phi = 0°$ during the first half cycle, the non-switching spins oscillate back through the wide potential minimum that is extended by the shoulder at $\phi = -115°$. This results in a strong red-shift of the centre frequency to 50 GHz. On the potential shoulder, the projection $\mathbf{F} \cdot \mathbf{k}_{NIR}$ drops below its initial value at $\phi_0$ (Extended Data Figure 5), leading to a transient negative offset of the Faraday signal (Fig. 4e, dashed-dotted line, and Extended Data Figure 6) until the oscillations of the unswitched spins have decayed within the starting local potential minimum. Still, a sufficiently large fraction of spins reach the target valley (grey sphere) for beating to be observed. Finally, α = 95° sets a new starting position and direction of acceleration (Fig. 4d, violet sphere and arrow), causing a reversal of the transient polarisation rotation signal and offset (Fig. 4e and Extended Data Figure 5). The wide potential minimum leads to a reduced centre frequency reproduced by calculating the single spin dynamics (Fig. 4f, black arrows). The large barrier to the neighbouring valley (grey circle) inhibits switching completely and no beating is observed.

The unprecedented phase slip, the asymmetric spectral splitting, and the long-lived offset in the magneto-optical response occurring above a well-defined threshold peak field are the fingerprints of ballistic spin switching, marking a novel regime of ultrafast all-coherent spin control throughout the entire phase space. In our specific implementation of a THz-driven anisotropy torque, the absorption of approximately one THz photon energy per spin suffices for switching whereas the energy dissipation within the spin system remains below 1 μeV per spin (see Methods). This scheme is, thus, highly scalable. Future storage devices could exploit the excellent spatial definition of antenna structures (Extended Data Figure 7) to switch magnetic bits of a diameter of 10 nm with THz energies of less than 1 attojoule. Owing to the absence of magnetic stray fields, these cells could be densely packed, similar to vortex core structures in ferromagnetic thin films[14]. The readout of the spin state could be combined with spintronic approaches[20,24]. Such optimised antennas with nanoscale gaps providing field enhancement factors of $10^4$ and more may be driven by all-electronic on-chip THz sources, enabling practical implementations of novel spin memories operating at THz clock rates, and ultimately low dissipation.

\* These authors contributed equally to this work.

§ Present address: Department of Physics, Lancaster University, Bailrigg, Lancaster, LA1 4YB, UK.

**Acknowledgement.** The authors thank A. M. Balbashov for bulk crystals of orthoferrites of an exceptionally high quality, J. Fabian and M. S. Sherwin for fruitful discussions and Th. Rasing for continuous support. The work in Regensburg was supported by the DFG through grant no. HU 1598/2 and SFB 1277 (Project A05) as well as by the European Research Council through grant no. 305003 (QUANTUMsubCYCLE). The work in Nijmegen was supported by the European Research Council ERC Grant Agreement No. 339813 (Exchange) and NWO (The Netherlands Organization for Scientific Research).

**Author Contributions** S.S., C.L., S.B., and R.H. designed and implemented the antenna structures. R.V.M. and A.V.K. identified the bulk material for the project. S.S., C.L., S.B., T.E., C.P.S. and D.C.V. carried out the experiment with support from R.V.M.. The theoretical modelling was carried out by C.L., S.S., S.B., A.K.Z., and R.V.M.. A.V.K. and R.H. supervised the study. All authors analysed the data, discussed the results, and contributed to the writing of the manuscript.

**Author Information** Reprints and permissions information is available at www.nature.com/reprints. The authors declare no competing financial interests. Correspondence and requests for materials should be addressed to C.L. (christoph.lange@ur.de) or R.V.M. (r.mikhaylovskiy@lancaster.ac.uk).

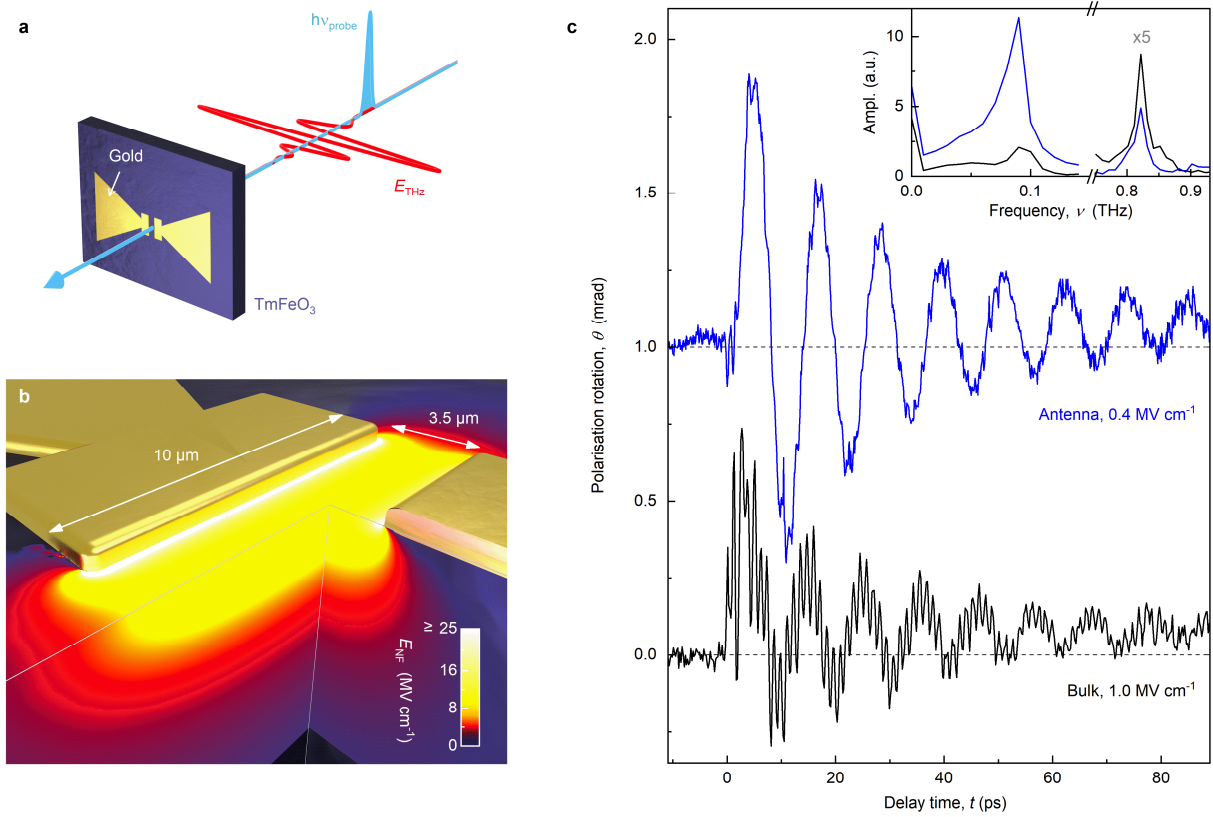

**Figure 1 | Antenna-enhanced THz spin dynamics. a,** Schematic of the gold bowtie antenna on TmFeO$_3$. The structure is excited from the back side by an intense THz electric field $E_{THz}$ (red waveform) while a co-propagating near-infrared pulse (h$\nu_{probe}$, light blue) probes the induced magnetisation dynamics in the centre of the feed gap. **b,** Peak near-field amplitude, $E_{NF}$, in the antenna feed gap calculated by finite-difference simulations for a real THz waveform with a peak field amplitude of $E_{THz}$ = 1.0 MV cm$^{-1}$ (see Extended Data Fig. 1c). **c,** Experimentally detected polarisation rotation signal as a function of the delay time, $t$, obtained for a peak electric THz field of $E_{THz}$ = 1.0 MV cm$^{-1}$ on the unstructured substrate (black curve) and when probing the gap of the bowtie antenna structure resonantly exited by a THz waveform with a peak electric far-field amplitude of $E_{THz}$ = 0.4 MV cm$^{-1}$ (blue curve, vertically offset by 1 mrad for better visibility). Inset: Corresponding amplitude spectra featuring two modes at 0.09 THz and 0.82 THz. The sample was kept at a lattice temperature of $T$ = 83 K.

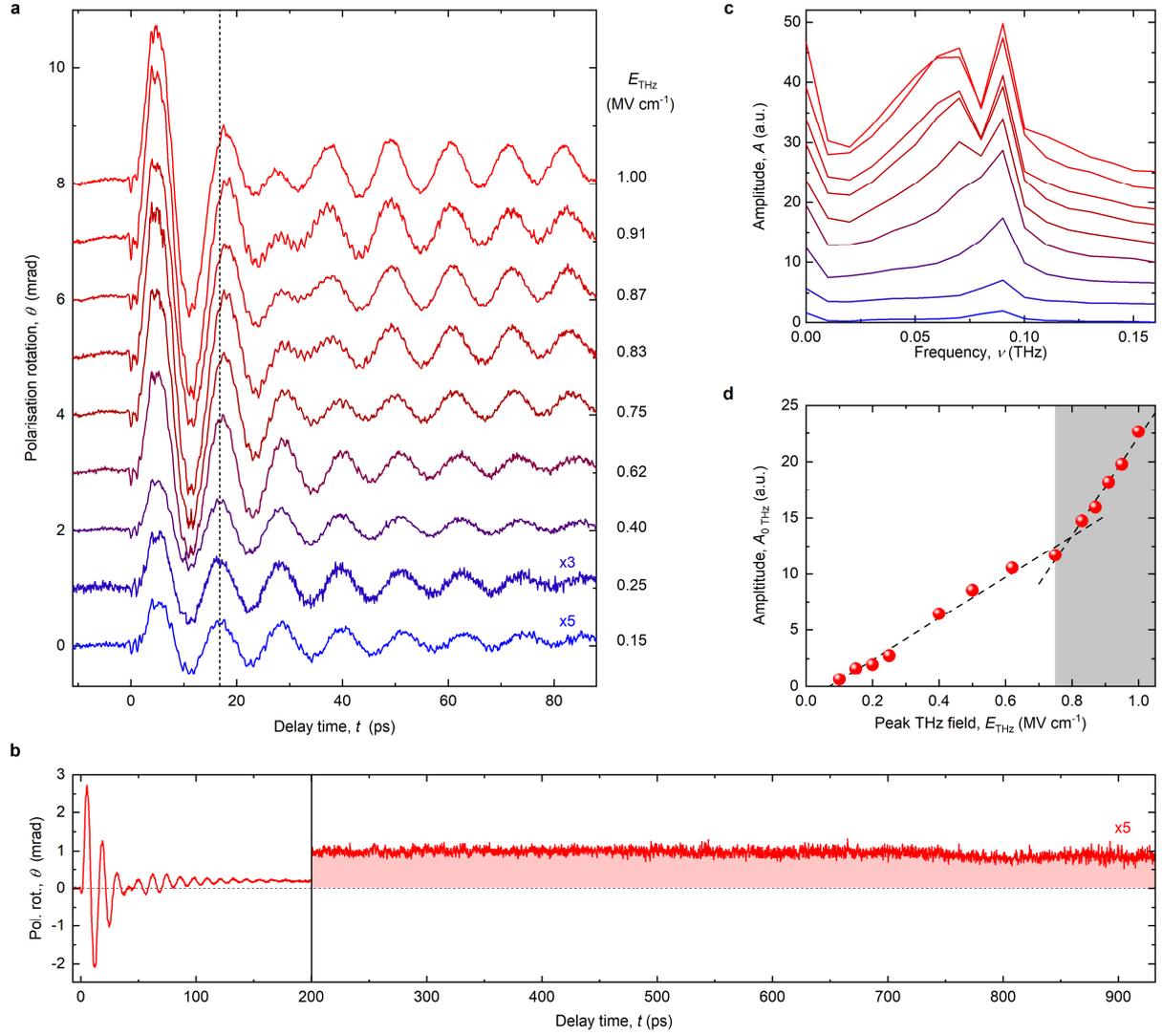

**Figure 2 | THz-induced nonlinear spin dynamics. a,** Polarisation rotation probed in the centre of the antenna feed gap for various far-field amplitudes, as a function of the delay time, $t$. For incident THz peak fields $E_{THz} >$ 0.75 MV cm$^{-1}$, the quasi-monochromatic oscillation is strongly distorted by a phase slip at delay times between 25 and 35 ps. The measurements are offset and scaled as indicated for clarity. Lattice temperature $T$ = 83 K. **b,** Long-term evolution of the polarisation rotation for a THz peak field of $E_{THz}$ = 1.0 MV cm$^{-1}$. The red-shaded area indicates the long-lived offset. **c,** Spectral amplitude of the time-domain data shown in **a**. The phase slip in the polarisation rotation signal for highest THz fields manifests itself in a splitting of the q-fm resonance. **d,** Spectral amplitude of the dc offset, $A_{0\,THz}$, as a function of the THz far-field peak amplitude, $E_{THz}$. $A_{0\,THz}$ increases monotonically with the THz field. Grey-shaded area: Spin-switching regime with increased slope of $A_{0\,THz}$. Dashed lines, guides to the eye.

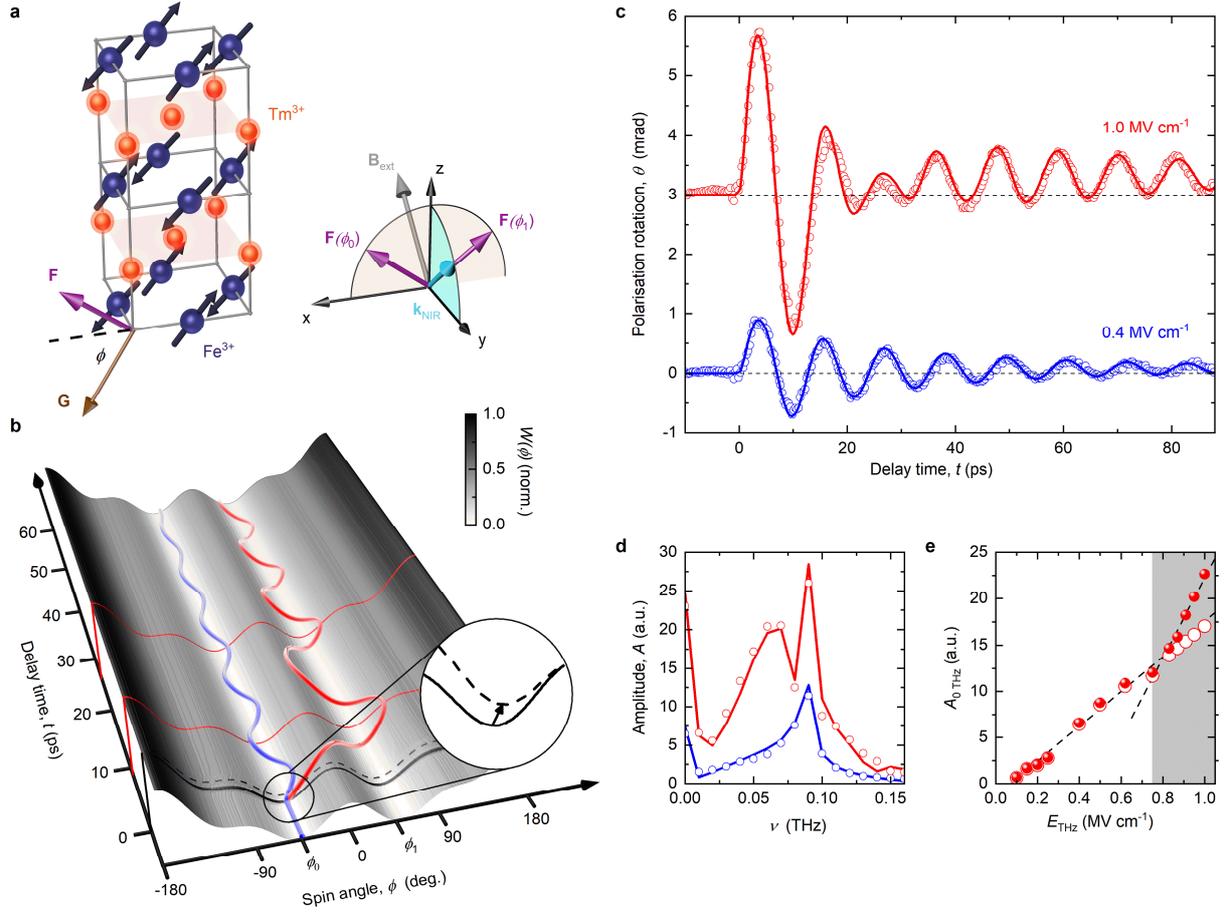

**Figure 3 | Microscopic picture of ballistic spin motion. a,** Spin and lattice structure of $TmFeO_3$ in the $\Gamma_{24}$ phase ($T_1 < T < T_2$), showing the $Fe^{3+}$ spins (dark blue spheres and arrows), $Tm^{3+}$ ions (orange spheres), and the ferromagnetic moment, **F** (violet arrow). The antiferromagnetic vector **G** (brown arrow) lies in the *x-z*-plane and encloses a finite angle of $0 < \phi < 90°$ with the *x*-axis. Inset: geometry of the wave vector of the probe pulse, $\mathbf{k}_{NIR}$ (light blue arrow), with respect to **F** and the external magnetic field $\mathbf{B}_{ext}$ (grey arrow). **b,** Numerical simulation of THz-induced ballistic spin dynamics. Upon THz excitation, the magnetic potential $W(\phi)$ is abruptly modified near a delay time of $t = 0$ ps (magnified in inset). Near-field THz transients with peak amplitudes of $E_{NF} = 6$ MV cm$^{-1}$ abruptly induce large-amplitude spin oscillations within the same potential valley around the initial angle $\phi_0$ (blue trajectory). For a THz near-field of $E_{NF} = 10$ MV cm$^{-1}$, the spins reach the adjacent local minimum (red trajectory) at $\phi_1$, where $\phi_1 \approx \phi_0 + 90°$, accumulating a phase retardation relative to spins oscillating around $\phi_0$ (delay times $t = 9.7$ ps and 27.2 ps, respectively; red cuts through the magnetic potential). **c,** Calculated polarisation rotation in the antenna feed gap for an incident THz electric peak field amplitude of $E_{THz} = 0.4$ MV cm$^{-1}$ (blue curve) and $E_{THz} = 1.0$ MV cm$^{-1}$ (red curve) as a function of the delay time, *t*, for a lattice temperature of $T = 83$ K, normalized to the experimental peak value. The experimental data are plotted as circles. **d,** Amplitude spectra of the time-domain data shown in **c**. **e,** Calculated scaling of the spectral amplitude of the long-lived offset, $A_{0\,THz}$, for no

misalignment (red circles) and a misalignment angle of the near-infrared **k**-vector out of the *y*-*z*-plane of 1.25° (red spheres). In the spin-switching regime ($E_{\text{THz}} \geq 0.75$ MV cm$^{-1}$, grey-shaded area) the calculations reproduce the increased slope of A$_{0\,\text{THz}}$ observed in the experiment (Fig. 2d). Dashed lines, guides to the eye.

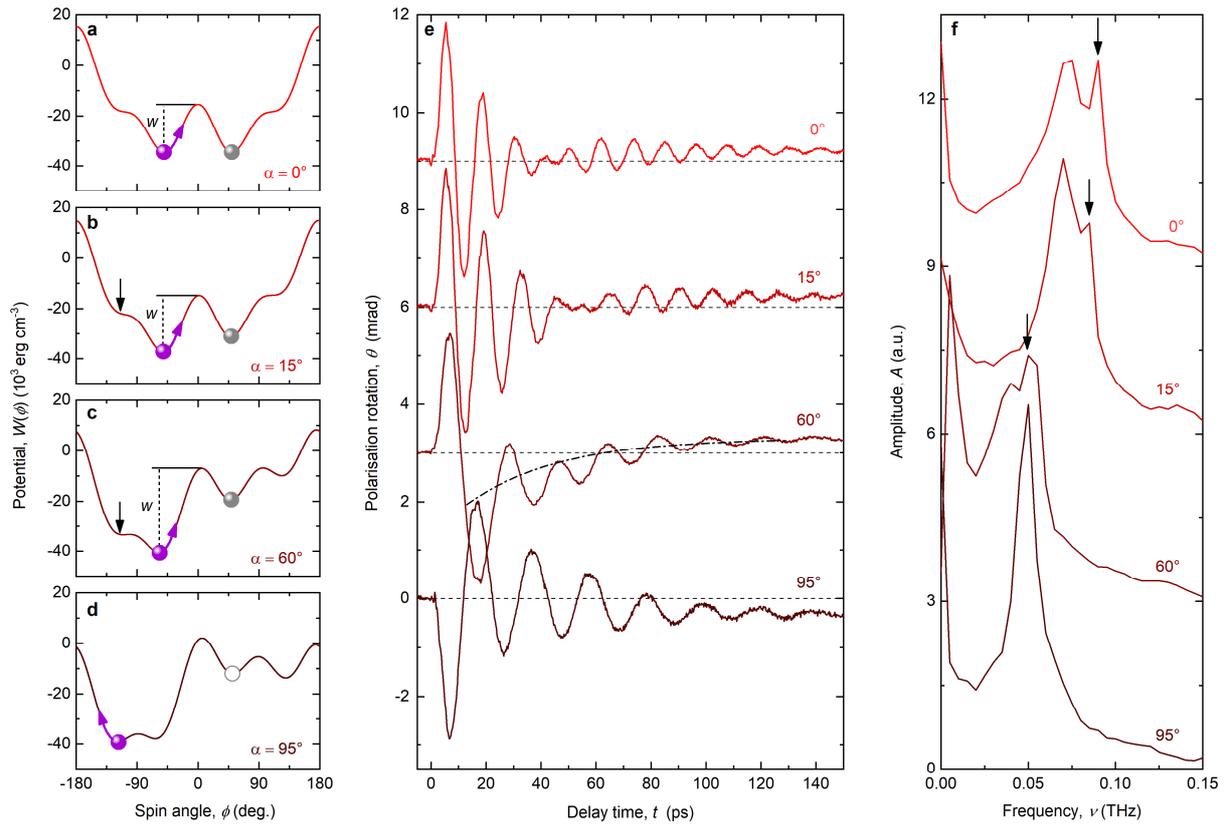

**Figure 4 | Ballistic navigation of spins. a-d,** Magnetic potential $W(\phi)$ for a lattice temperature of $T$ = 82.5 K and various orientations α of the static external magnetic bias, $\mathbf{B}_{ext}$. $w$, potential barrier height relevant for switching; black arrows, potential shoulder associated with the red-shift. Violet spheres and arrows: initial spin orientation and direction after excitation; grey spheres, final orientation of switched spins. **e,** Polarisation rotation as a function of the delay time, $t$, for the potentials shown in **a-d** and a THz peak far-field amplitude of $E_{THz}$ = 1.0 MV cm$^{-1}$. Dashed-dotted curve: transient negative polarisation rotation (see text). **f,** Amplitude spectra of the time-domain data shown in **e**. The black arrows mark the computed centre frequencies.

**Methods**

**Sample preparation.** We used a monocrystalline, 60-µm-thick TmFeO$_3$ sample obtained by floating-zone melting. The sample was cut perpendicularly to one of the crystal's optical axes, which lies in the *y-z*-plane at an angle of 51° with respect to the *z*-axis. The custom-tailored THz antennas with a feed gap of 3.5 µm and a resonance frequency of 0.65 THz were processed on top of the crystal by electron-beam lithography of a poly(α-methylstyrene-co-α-chloracrylate methylester) resist, subsequent evaporation of 100 nm of gold, and lift-off. The structure was kept in a helium cryostat and cooled to temperatures within the $\Gamma_{24}$ transition phase. For the measurements discussed in the first part of the manuscript, a static bias field of $\mathbf{B}_{ext}$ = 100 mT from a permanent magnet was applied within the *y-z*-plane of the crystal at an angle of 39° relative to the *z*-axis, defining the equilibrium spin orientation $\phi_0$ and ensuring the restoring of the magnetisation between subsequent laser pulses. For the data shown in Fig. 4, the **B**-field was rotated about the optical axis of the crystal, whereby an angle of α = 0° denotes the starting position within the *y-z*-plane as defined above.

**Experimental setup.** Intense single-cycle THz pulses were generated by tilted-pulse front optical rectification of near-infrared pulses from a low-noise Ti:sapphire laser amplifier (centre wavelength, 807 nm; pulse energy, 5.5 mJ; pulse duration, 33 fs; repetition rate, 3 kHz) in a cryogenically cooled LiNbO$_3$ crystal (Extended Data Figure 1b). A pair of wire-grid polarisers were used to control the peak field strength and the polarisation state of the THz waveforms. Extended Data Figure 1c and d show the THz transient and the corresponding spectrum featuring frequency components between 0.3 and 2.5 THz. A small portion of the near-infrared power was sent through a delay line, combined with the THz pulse using a fused silica window coated with indium tin oxide, and collinearly transmitted through the feed gap of the antenna structure to probe the magnetisation state. The polarisation rotation was measured by subsequent optics consisting of a half-wave plate, a Wollaston prism, and two balanced silicon photodiodes, read out by a lock-in amplifier. In order to calibrate the THz field amplitude in the sample focus, the focal diameter was determined by measuring the THz power transmitted through a circular aperture. The corresponding total energy was then calculated by time-integration of $E_{THz}(t)^2$. We estimate that the total uncertainty of the field amplitude amounts to 16%.

**Estimate of the spin switching energy.** The Poynting theorem dictates that the absorbed electromagnetic power density $P(t)$ is given by

$$P(t) = j(t) E(t), \qquad (1)$$

where $j(t)$ is the effective current density describing dissipative processes in a material and $E(t)$ is the oscillating electric field. The full energy absorbed per unit volume is therefore

$$W_{\text{abs}} = \int_{-\infty}^{\infty} j(t) E(t) \, dt. \qquad (2)$$

By taking the Fourier transforms $j(t) = \frac{1}{2\pi}\int_{-\infty}^{\infty} \tilde{j}(\omega) e^{i\omega t} \, d\omega$ and $E(t) = \frac{1}{2\pi}\int_{-\infty}^{\infty} \tilde{E}(\omega') e^{i\omega' t} \, d\omega'$, where $\omega$ is the frequency, and substituting them into Eq. (2) we obtain

$$W_{\text{abs}} = \frac{1}{2\pi} \int_{-\infty}^{\infty} \tilde{j}(\omega) \tilde{E}(-\omega) \, d\omega. \qquad (3)$$

The current density is connected to the electric field by the effective conductivity $\sigma(\omega) = \frac{\tilde{j}(\omega)}{\tilde{E}(\omega)}$ so as

$$W_{\text{abs}} = \frac{1}{2\pi} \int_{-\infty}^{\infty} \sigma(\omega) \tilde{E}(\omega) \tilde{E}(-\omega) \, d\omega = \frac{1}{2\pi} \int_{-\infty}^{\infty} \sigma(\omega) |\tilde{E}(\omega)|^2 \, d\omega. \qquad (4)$$

In the case of crystal-field split ground state transitions of $TmFeO_3$ in the temperature interval between 80 K and 90 K, where the imaginary part of the dielectric function $\varepsilon_2$ is much smaller than its real part $\varepsilon_1$ (see Ref. 31), the effective conductivity can be approximated by $\sigma = \varepsilon_0 \, n_{\text{sub}} c \, \alpha_{\text{eff}}$. Here $n_{\text{sub}} = 4.92$ is the refractive index of $TmFeO_3$, and $\alpha_{\text{eff}} \approx 4000 \, \text{m}^{-1}$ is the effective THz absorption coefficient obtained from data of Ref. 31, taking into account the spectral shape of our THz pulse. We obtain

$$W_{\text{abs}} = \frac{1}{2\pi} \varepsilon_0 \, n_{\text{sub}} \, c \, \alpha_{\text{eff}} \int_{-\infty}^{\infty} |\tilde{E}(\omega)|^2 \, d\omega, \qquad (5)$$

which can be rewritten in the time domain (compare Eqs. (2) and (3)) as

$$W_{\text{abs}} = \varepsilon_0 \, n_{\text{sub}} \, c \, \alpha_{\text{eff}} \int_{-\infty}^{\infty} E^2(t) \, dt. \qquad (6)$$

The absorbed energy density in the rare-earth system for a near-field THz transient with a peak electric field of 7.8 MV cm$^{-1}$, which exceeds the threshold for spin switching, is $W_{\text{abs}} = 20$ J cm$^{-3}$. $TmFeO_3$ crystallises in a distorted perovskite structure with a unit cell volume of $V_{\text{uc}} = 2.22 \times 10^{-28}$ m$^3$ (lattice constants, $a = 525$ pm, $b = 557$ pm, and $c = 758$ pm) (see Ref. 32), which contains 4 $Fe^{3+}$ spins. Thus, an upper bound for the absorbed energy in the rare-earth system per spin is given by $W_{\text{spin}} = W_{\text{abs}} \times \frac{V_{\text{uc}}}{4} = 7.15$ meV, which is on the order of the energy of one THz photon. The dissipation by the spin system is even smaller: The energy required to overcome the potential barrier, separating two neighbouring potential minima (see Fig. 3b), normalized by the number of spins in the switched volume is less than 1 µeV. This value can, thus, be regarded as an upper limit for the maximal energy dissipated by one spin upon switching.

**Estimate of the magnetisation deflection in the near-field volume.** In the case of unstructured bulk TmFeO$_3$, the total polarisation rotation, $\theta$, results from approximately equal contributions across the entire sample thickness of 60 µm. In order to calibrate the relation between $\theta$ and the spin angle $\phi$, we enforce a full switching of the magnetisation (change of $\phi$ by 180°) by reversing the external static magnetic bias field. This scenario rotates the probe polarisation by 24 mrad. Thus, we conclude that a polarisation rotation of $\theta = 0.5$ mrad, as induced by a THz amplitude of 1.0 MV cm$^{-1}$ in the antenna-free sample, corresponds to a transient spin excursion of $\Delta\phi = 3.5°$. Taking into account the quadratic dependence of $\Delta\phi$ on the electric field amplitude[26], we link the polarisation rotation to the THz peak electric field by $\theta = \xi \times L \times \bar{E}_{\text{peak}}^2$, where $L$ is the crystal length, $\xi = 472$ mrad cm (MV)$^{-2}$ is the coupling constant, and $\bar{E}_{\text{peak}} = 0.42$ MV cm$^{-1}$ is the peak electric THz amplitude averaged over the length of the unstructured TmFeO$_3$ sample. In the antenna-covered structure, the magneto-optical signal can be divided into two contributions: the antenna near-field region extending down to a depth of 13 µm below the antenna (Extended Data Figure 8, red-shaded area), where electric fields strongly exceeding the far-field amplitude are encountered, and a bulk part (Extended Data Figure 8, blue-shaded area), where the electric field assumes an average value of 0.3 MV cm$^{-1}$. Accordingly, the polarisation rotation by the bulk part is $\theta_b = \xi \times 47$ µm $\times (0.3$ MV cm$^{-1})^2 = 0.2$ mrad, such that 0.7 mrad of the total magneto-optical signal result from the near-field volume. This contribution corresponds to an average spin deflection angle of $\Delta\phi = 24°$.

**Numerical calculation of antenna response.** The THz response of the entire structure, including the near-field of the custom-tailored antenna as well as the substrate, was obtained by solving Maxwell's equations using a finite-difference frequency-domain (FDFD) approach. The refractive index of TmFeO$_3$ is set to $n_{\text{sub}}$ = 4.92, while the gold structure is implemented as a perfect metal. The THz near-field waveforms were subsequently calculated based on the measured far-field THz waveform, employing the results of the FDFD calculations as a complex-valued transfer function. These near-field waveforms enabled us to retrieve the local dynamics of the spin deflection angle, $\phi$, by time-domain numerical integration as detailed below. The overall polarisation rotation was obtained by integrating the local contributions along the entire probe volume, weighed by the intensity profile of the probe beam. We used a diameter of 6 µm (FWHM) in the direction parallel to the capacitor plates, and 2 µm (FWHM) in the orthogonal direction in order to account for diffraction effects near the capacitive plates. While calibrating near-fields in excess of ~10 MV cm$^{-1}$ is challenging[27,28], the total polarisation rotation is robust against variations of the maximum near-fields occurring only in the close vicinity of the capacitive plates, as confirmed by calculations. A grid resolution of (100 nm)$^3$ was chosen for proper convergence.

**Calculation of spin dynamics.** We adapted the previously derived formalism for THz-induced spin dynamics based on the generalized sine-Gordon equations for our high-field setting[26]. The vectorial spin orientation can be mapped onto the angle $\phi$ between the antiferromagnetic vector **G** and the *x*-axis (Fig. 3a). The magnetic potential $W(\phi)$ of TmFeO$_3$ is given by[26]

$$W(\phi) = K_1 \sin^2 \phi + K_2 \sin^4 \phi - \frac{H_D}{H_E} M_{Fe}(B_{\text{ext}} \cos \alpha \cos \phi - B_{\text{ext}} \sin \alpha \, \sin \phi - B_{\text{THz}} \sin \phi),$$

(7)

where $H_D$ = 2 × 10$^5$ Oe is the Dzyaloshinskii field, $H_E$ = 2 × 10$^7$ Oe is the effective field of the *d-d* exchange, and $M_{Fe}$ = 1000 e.m.u. cm$^{-3}$ is the magnetisation of a single Fe$^{3+}$ sublattice[33]. The parameter $K_1 = 2K_2 \frac{T-T_2}{T_2-T_1}$ for $T_1 < T < T_2$, where $K_2$ is a constant, sets the potential curvature by the frequency of the quasi-ferromagnetic mode $\omega_{\text{q-fm}}^2 = \frac{1}{2} \omega_E \omega_A \sin^2 \phi_0$ in the linear regime of spin dynamics. Here, $\omega_E = \gamma H_E$, $\omega_A = \gamma \frac{K_2}{M_{Fe}}$, $\gamma$ is the gyromagnetic ratio, $T$ is the spin lattice temperature, and $T_1$ = 80 K and $T_2$ = 90 K are the lower and upper temperature bounds of the $\Gamma_{24}$ transition phase, respectively. The thermal excitations of the crystal-field-split ground states determine the equilibrium angle of the spin vector, $\phi_0 = \arcsin\left(\frac{T-T_2}{T_2-T_1}\right)^{\frac{1}{2}}$ (see Ref. 26). For our numerical simulations, we calibrated the effective magnetic potential $W(\phi)$ by the experiment with bulk TmFeO$_3$, and we included an external magnetic field along the $z$-axis ($\alpha = 0$) of B$_{ext}$ = 150 mT compatible with the experimentally determined value. As we are operating in the high-field regime, where the THz-induced nonlinear anisotropy torque dominates[26], we neglect the magnetic THz spin interaction with the THz magnetic field, $B_{THz}$, which is oriented along the crystallographic $x$-axis.

The equation of motion accounting for a THz-induced change of the magnetic potential energy reads

$$\ddot{\phi} - C^2 \nabla^2 \phi = -\gamma_D \dot{\phi} + \omega_E \omega_A \cos(\phi) \sin(\phi) \times (\eta + \sin^2(\phi)) +$$

$$\kappa \cos(\phi) \sin(\phi)\, \varepsilon_0\, n_{sub}\, c\, \alpha_{eff} E_{THz}^2 - \frac{H_D}{H_E} \gamma\, w_E\, B_{ext} \sin\phi. \qquad (8)$$

Here, $\gamma_D$ is the damping. The excitation by the crystal field transitions is modelled by both an impulsive and a displacive mechanism, accounting for an increase of the angular velocity, $\dot{\phi}$, and a shift of the equilibrium spin angle, $\phi_0$, respectively, in conceptual analogy to Ref. 34. The impulsive excitation is implemented by the term proportional to the constant $\kappa$, coupling the spin dynamics to the instantaneous THz power density $\varepsilon_0 n_{sub} c \alpha_{eff} E_{THz}^2$. To account for the displacive term, we implement a strong THz-induced excitation of the crystal field transitions, leading to an increase of the population density $\Delta\rho(t)$ of the excited states of the Tm$^{3+}$ ions. In our model, this is described by the excitation parameter $\eta = \frac{(\rho(T)+\Delta\rho(t))-\rho_2}{\rho_2-\rho_1}$, where $\rho(T)$, $\rho_1$, and $\rho_2$ are the equilibrium population densities of the crystal-field split states at the temperature $T$, $T_1$, and $T_2$, respectively. The THz-induced change of the population density leads to an abrupt change of the magnetic potential, $W(\phi)$, of the iron spins, resulting in a displacive anisotropy torque. Quantitatively, we calculate $\Delta\rho(t) = \Gamma \int_{-\infty}^{t} \frac{\varepsilon_0\, n_{sub}\, c\, \alpha_{eff}}{\hbar \omega_{CFT}} E_{THz}^2(t')\, dt'$, where $\Gamma$ is a coupling parameter, $\hbar$ is Planck's constant, and $\omega_{CFT}$ is the resonance frequency of the electric dipole active Tm$^{3+}$ ground state transition[35]. The term $C^2 \nabla^2 \phi$ accounts for the interaction between different magnetic domains of the sample, where $C$ is the spin wave velocity that sets the maximal speed of a domain boundary. In the orthoferrites, $C = 2 \times 10^6$ cm s$^{-1}$ (see Ref. 36, 37). One can see that, on the ~1 ps timescale of our experiment, the regions of the sample exposed to the THz fields of different strengths can be assumed to be practically non-interacting as the magnetic excitations travel a distance of 10 nm during this time. This distance is also much smaller than the characteristic spatial scale of the THz near-field of >1 µm. We therefore neglected the term $C^2\nabla^2\phi$ in our numerical simulations.

The local dynamics of the spin deflection angle, $\phi$, are calculated by solving equation (8) separately for each near-field cell using the corresponding THz near-field transient (see Supplementary Video 1). As confirmed by polarimetry, the THz-induced change of the magnetisation leads to a rotation of the near-infrared probe polarisation. A switch-off analysis shows that the Faraday rotation is almost exclusively caused by the ferromagnetic component of the magnetisation, while the dynamics of the antiferromagnetic response plays a minor role. Thus, the microscopic Faraday rotation is obtained by projecting the ferromagnetic vector, $\mathbf{F}(\phi)$, of each cell onto the wave vector of the near-infrared probe beam, $\mathbf{k}_{NIR}$. Integration of these contributions along the optical axis allows us to quantitatively reproduce the experimentally detected polarisation rotation, $\theta$, (see Fig. 3c). In the non-perturbative regime, the actual spin trajectory depends sensitively on the exact location within the near-field region of the antenna. Yet the total magneto-optical response integrated over the entire near-field volume is fairly robust against minor field fluctuations. For our measurement with a far-field THz peak amplitude of $E_{THz} = 0.4$ MV cm$^{-1}$, we obtain the best agreement (Fig. 3c, blue curve) using the experimentally determined spin dephasing rate $\gamma_D = 45$ GHz, as well as the following values: $\omega_{q-fm}/2\pi = 88.7$ GHz, $\kappa = 3.58 \times 10^8$ m$^2$ Ws$^{-2}$, and $\Gamma = 2.49 \times 10^{-10}$ m$^3$s. For a THz peak amplitude of $E_{THz} = 1.0$ MV cm$^{-1}$ (Fig. 3c, red curve), we slightly adjust some of the parameters to $\omega_{q-fm}/2\pi = 90.0$ GHz, $\kappa = 1.02 \times 10^8$ m$^2$ Ws$^{-2}$, and $\Gamma = 1.01 \times 10^{-10}$ m$^3$s. Magnon-magnon scattering can effectively be accounted for by introducing a momentum dependent damping in the spin system. Extended Data Figure 9 shows the results of a switch-off analysis considering three scenarios including the full calculation (solid lines), only the displacive (dashed lines), and only the impulsive contribution (dashed-dotted lines). Whereas for a field amplitude of $E_{THz} = 0.4$ MV cm$^{-1}$, the sum of displacive and impulsive contributions approximates the full calculation, the strong-field dynamics at $E_{THz} = 1.0$ MV cm$^{-1}$ are only rendered correctly by the full calculation. In all cases, a purely displacive effect yields an exclusively positive magneto-optical signal and a non-zero signal offset, while the impulsive component is responsible for the strong oscillatory component.

**Methods References**

**Data Availability.** The data supporting the findings of this study are available from the corresponding authors upon request.

**Supplementary Information**

**Supplementary Video 1 | Visualisation of calculated local spin dynamics in the antenna near-field.** Top panel, measured (grey circles) and calculated (red curve) polarisation rotation signal for $E_{THz} = 1.0$ MV cm$^{-1}$ (Fig. 3c, red curve). Lower set of panels, *x-y*-, *x-z*- and *y-z*-projections of the calculated spin dynamics in the antenna near-field as a function of the delay time, *t*.

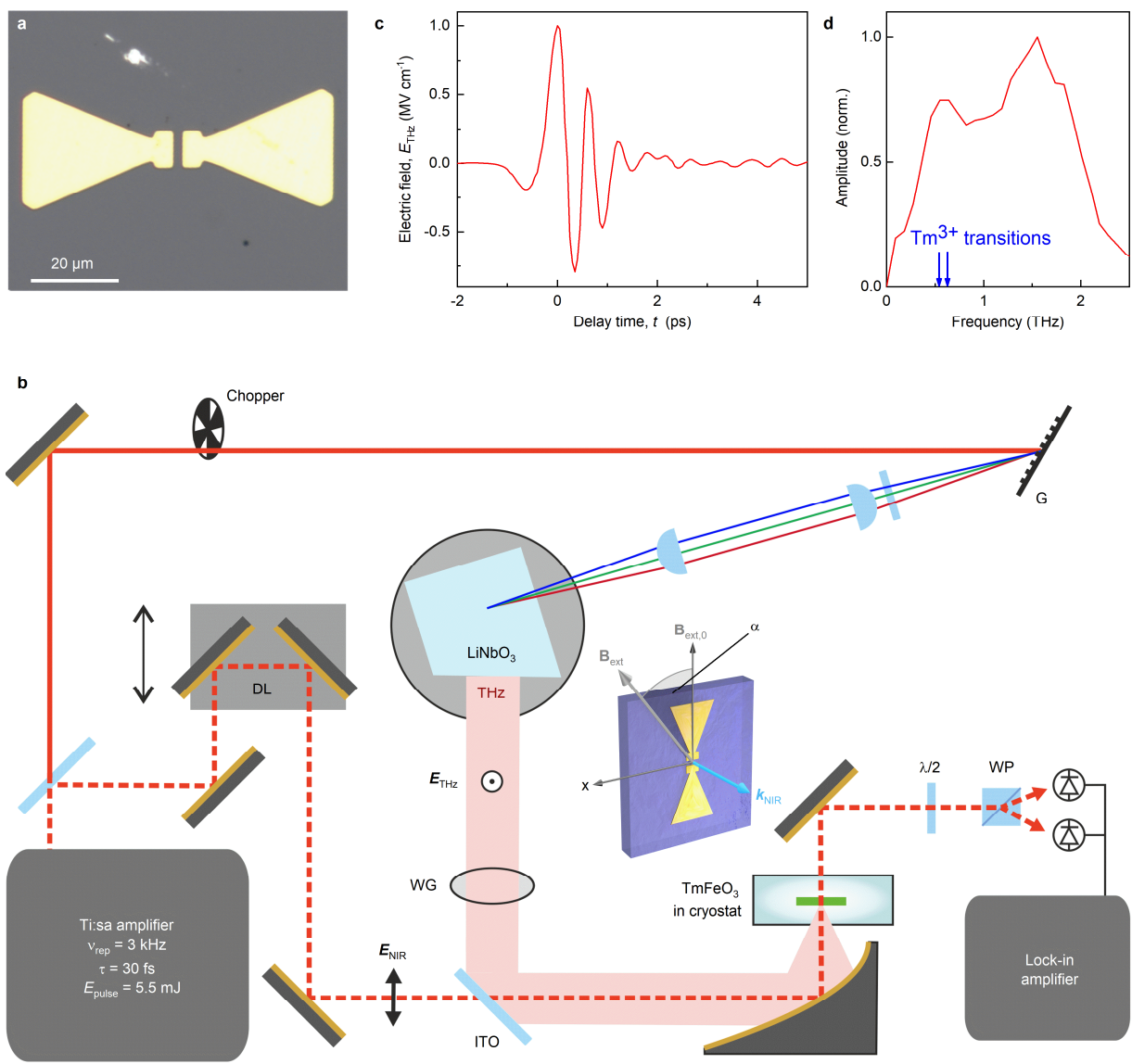

**Extended Data Figure 1 | Experimental setup. a,** Microscope image of the gold bowtie antenna with a resonance frequency of 0.65 THz and a feed gap of 3.5 µm, structured onto the TmFeO$_3$ sample. **b,** Ti:sapphire amplifier, centre wavelength, 807 nm; pulse energy, 5.5 mJ; pulse duration, 33 fs; repetition rate, 3 kHz. The grating (G), imprints a pulse front tilt onto the near-infrared beam. Two cylindrical lenses image and focus the laser light into a cryogenically cooled lithium niobate crystal (LiNbO$_3$). WG, pair of wire grid polarisers controlling the intensity and the polarisation state of the generated THz pulses. ITO, indium tin oxide coated calcium fluoride window. The THz-induced polarisation changes are decoded with the help of a half-wave plate (λ/2), a Wollaston polariser (WP) and a pair of photodiodes and subsequently detected with a lock-in amplifier. DL, mechanical delay line. $E_{NIR}$, near-infrared probe pulse polarisation. $E_{THz}$, THz polarisation. The inset depicts the orientation of the static magnetic field, $B_{ext}$, as a function of the angle α relative to the orientation $B_{ext,0}$ used for the measurements in the first part of the manuscript. **c,** Electro-optically detected THz field, $E_{THz}$, generated by tilted-pulse front optical

rectification. **d,** Corresponding spectral amplitude of the THz transient shown in **c**. The blue arrows indicate the frequencies of the $Tm^{3+}$ ground state transitions relevant for our experiment.

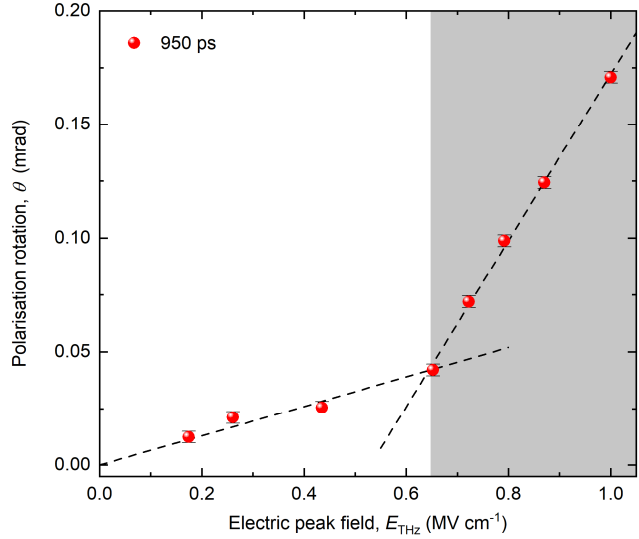

**Extended Data Figure 2 | Scaling of the residual offset for large delay times.** Polarisation rotation signal at a delay time of $t$ = 950 ps as a function of the THz electric peak field, $E_{THz}$. The data are extracted from time-resolved measurements in the feed gap of an antenna structurally similar to the one discussed in the main text with a feed gap of 3.5 µm and a broad resonance around 0.65 THz, optimised to the $Tm^{3+}$ ground state transitions. Lattice temperature $T$ = 81 K. In the spin switching regime, $E_{THz}$ > 0.65 MV cm$^{-1}$, the slope of the polarisation rotation signal is significantly increased. Error bars, standard deviation of $\theta$. Dashed lines, guides to the eye.

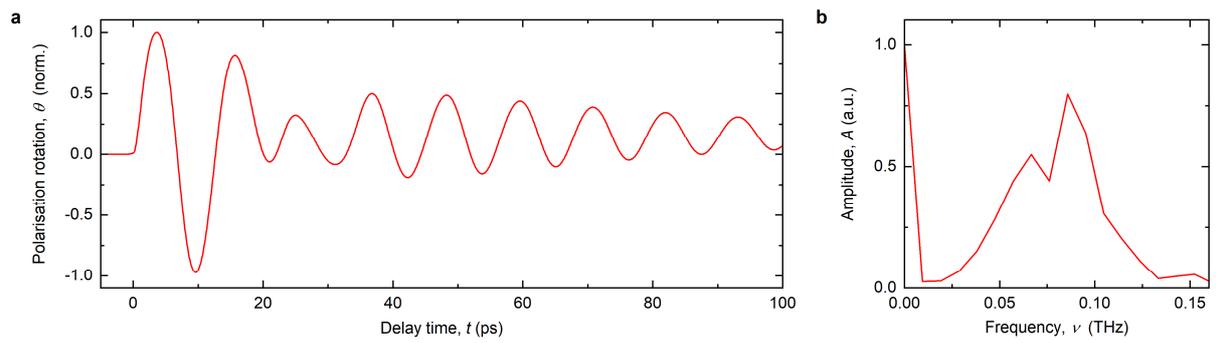

**Extended Data Figure 3 | Qualitative simulation of the beating signature. a,** Polarisation rotation calculated by superimposing the responses shown in Fig. 3b, that is, spins oscillating in the equilibrium potential minimum at $\phi_0$ (relative weight, 0.8) and spins driven into the neighbouring local minimum at $\phi_1$ (relative weight, 0.2). **b,** Amplitude spectra of the time-domain data shown in **a**.

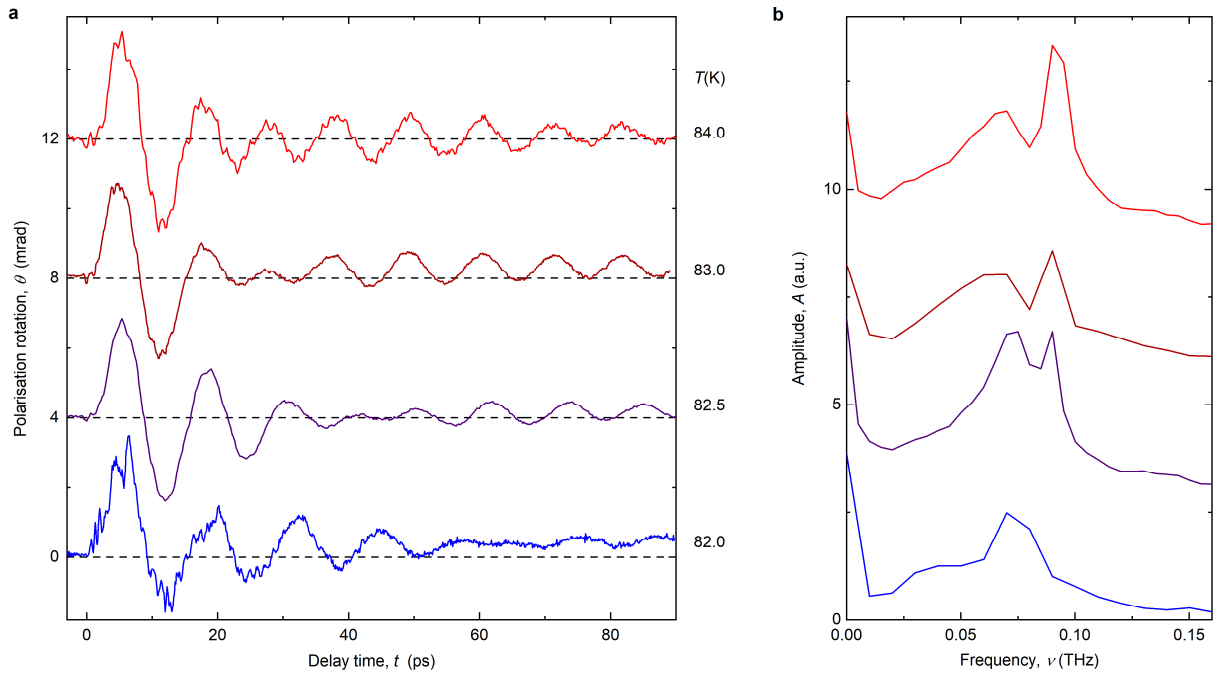

**Extended Data Figure 4 | Temperature dependence of spin dynamics. a,** Transient polarisation rotation probed in the centre of the feed gap of the antenna discussed in Fig. 4, for a THz far-field amplitude $E_{THz} = 1.0$ MV cm$^{-1}$ and different lattice temperatures, $T$, between 82.0 K and 84.0 K. **b,** Corresponding amplitude spectra of the data shown in **a**.

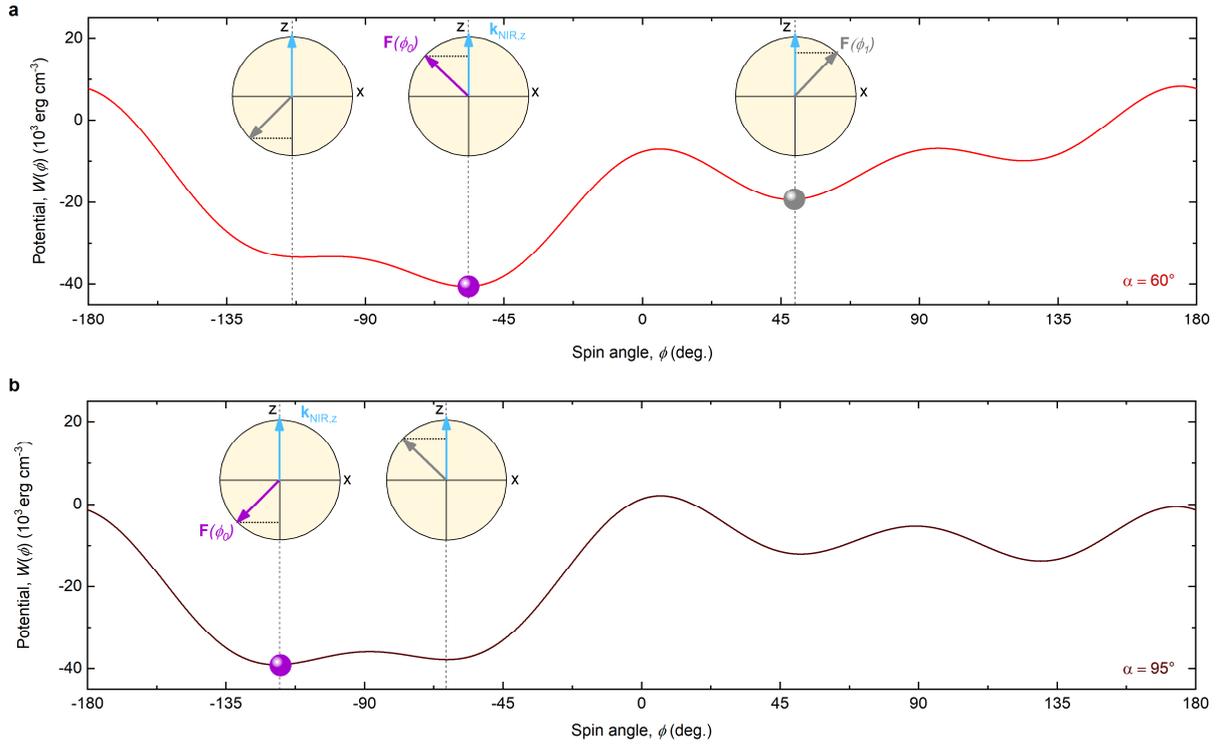

**Extended Data Figure 5 | Faraday signal for spin dynamics in different magnetic potentials. a,** Magnetic potential (red curve) for a lattice temperature of $T = 82.5$ K and an angle of $\mathbf{B}_{ext}$, $\alpha = 60°$, as shown in Fig. 4c. Violet (grey) sphere, initial (switched) spin state. Insets: projection (grey dotted horizontal lines) of the magnetisation $\mathbf{F}(\phi)$ (arrows) onto the near-infrared wave vector, $\mathbf{k}_{NIR,z}$ (light blue arrow), for different angles $\phi$. For $\phi < \phi_0$, the projection drops below its initial value and becomes negative for $\phi < -90°$, causing a negative transient Faraday signal (Fig. 4e). For $\phi_0 < \phi < \phi_1$, $\mathbf{k}_{NIR} \cdot \mathbf{F}(\phi) > \mathbf{k}_{NIR} \cdot \mathbf{F}(\phi_0)$, resulting in the positive initial half-cycle of the Faraday rotation signal (Fig. 4e). **b,** Magnetic potential for $\alpha = 95°$ (dark red curve) as shown in Fig. 4d. For $\phi < \phi_0$, the initial spin deflection leads to $\mathbf{k}_{NIR} \cdot \mathbf{F}(\phi) < \mathbf{k}_{NIR} \cdot \mathbf{F}(\phi_0)$, causing a negative onset of the first oscillation period (Fig. 4e, bottom curve).

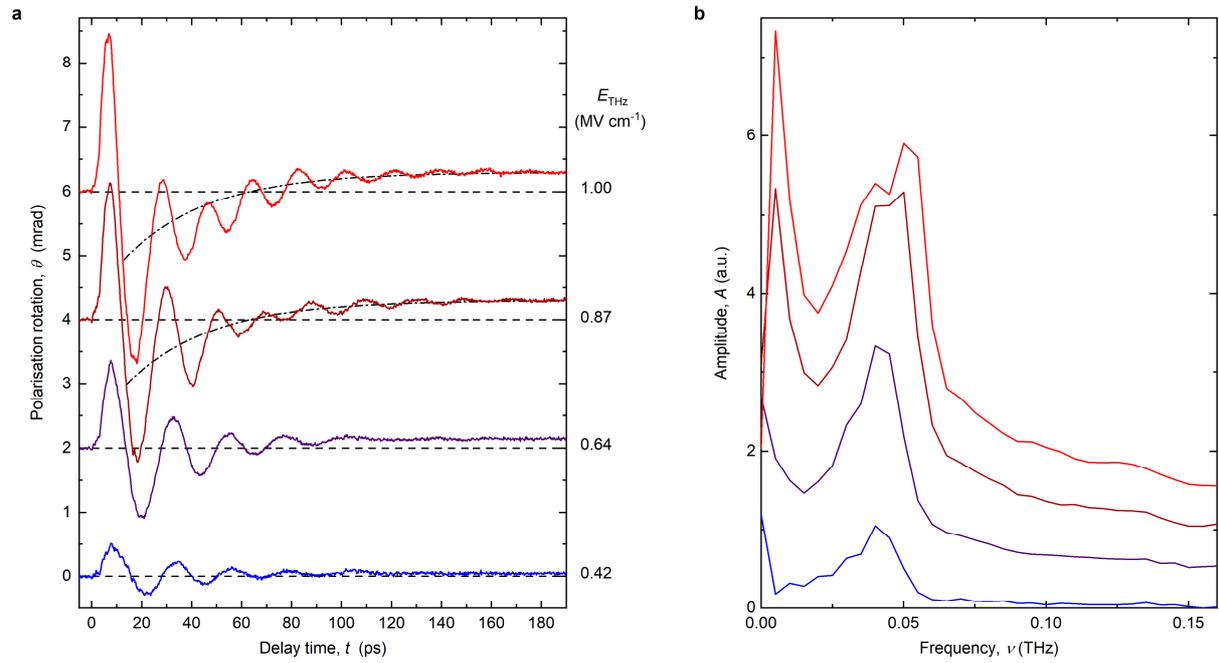

**Extended Data Figure 6 | Field dependence of spin dynamics for α = 60°. a**, Polarisation rotation signal as a function of the delay time, $t$, for different THz fields, $E_{THz}$, between 0.42 and 1.0 MV cm$^{-1}$, probed in the centre of the feed gap of the antenna discussed in Fig. 4. The transient negative Faraday signal (dashed-dotted curves) builds up for $E_{THz} \geq 0.87$ MV cm$^{-1}$. **b,** Corresponding amplitude spectra of the data shown in **a**.

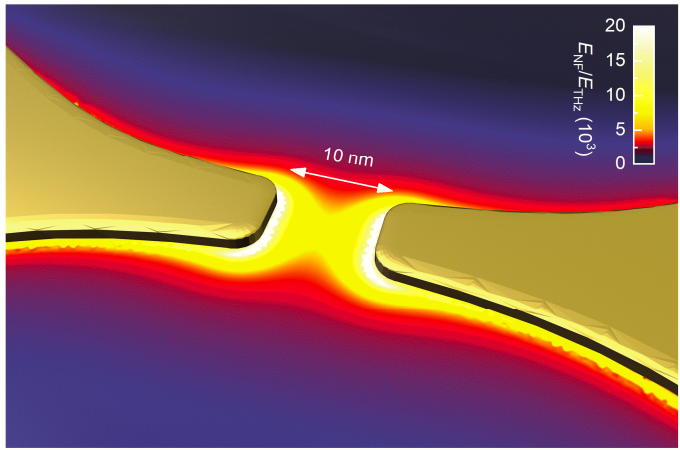

**Extended Data Figure 7 | Electric field enhancement in the near-field of a THz nanoantenna.** Enhancement factor $E_{NF}/E_{THz}$ of the near-field peak amplitude $E_{NF}$ compared to the THz electric far-field $E_{THz}$ calculated by finite-difference simulations for a real THz waveform in the near-field of an antenna structure with a feed gap of 10 nm. Assuming a switching threshold of ~10 MV cm$^{-1}$ a far-field amplitude of only 1 kV cm$^{-1}$ is sufficient to drive coherent spin switching by 90° in the centre of the antenna structure.

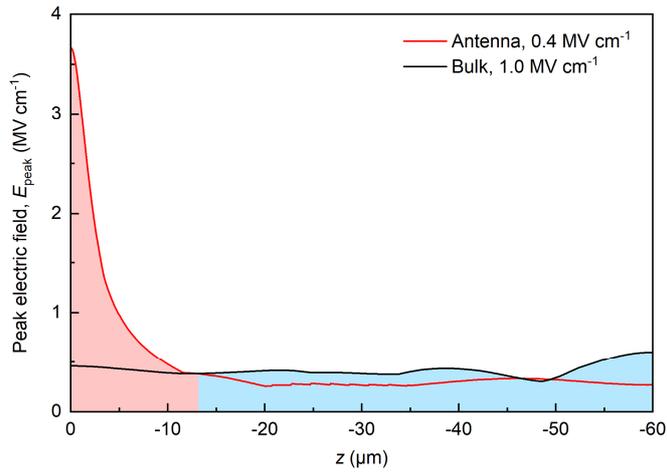

**Extended Data Figure 8 | Calculated electric near-field characteristics of antenna.** Near-field amplitude $E_{NF}$ as a function of depth $z$ in the center of the antenna feed gap, for a THz far-field amplitude of $E_{THz} = 0.4$ MV cm$^{-1}$ (red curve). The electric field distribution expected in the unstructured substrate, for $E_{THz} = 1.0$ MV cm$^{-1}$ is shown for comparison (black line). The near-field region of the antenna, where the electric field exceeds the value of the bulk structure, is indicated by the red-shaded area.

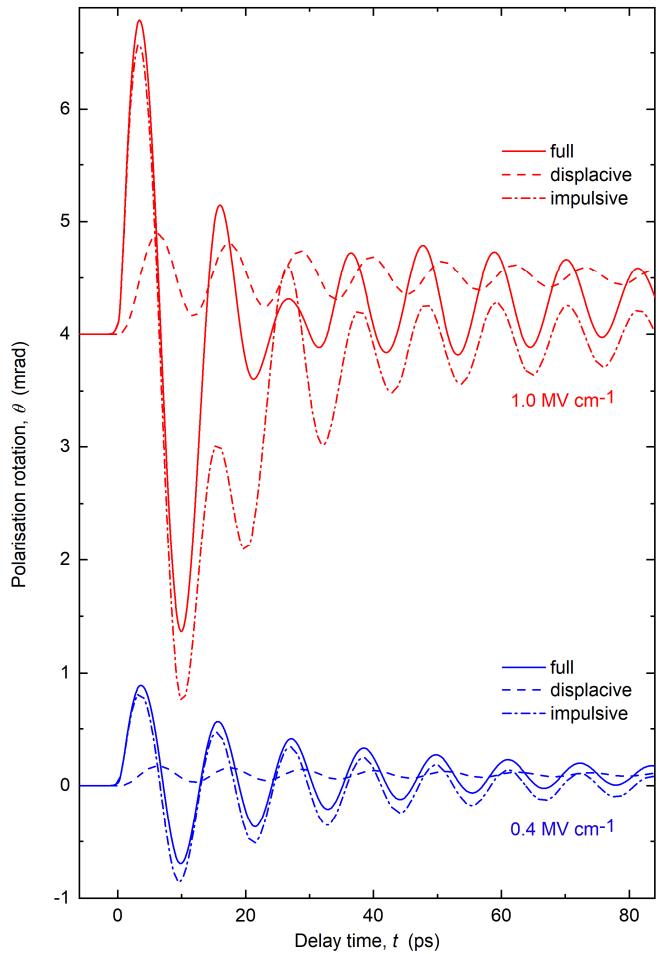

**Extended Data Figure 9 | Simulated magneto-optical response for different driving forces.** Calculated polarisation rotation signals expected from the antenna structures for a THz far-field amplitude of 0.4 MV cm$^{-1}$ (blue curves) and 1.0 MV cm$^{-1}$ (red curves). Calculations including only the displacive (dashed lines) or impulsive (dashed-dotted lines) anisotropy torque do not fit the experimental data. For the switch-off analysis, the parameters $\Gamma$ for the displacive and $\kappa$ for the impulsive torque of the full calculation (solid lines) are used. The curves are offset and normalized to the experimental peak value.